\documentclass[apjl]{emulateapj}

\begin{document}
\title{Detection of Extended VHE Gamma Ray Emission from G106.3+2.7 with VERITAS}

\author{
V.~A.~Acciari\altaffilmark{1},
E.~Aliu\altaffilmark{2,*},
T.~Arlen\altaffilmark{3},
T.~Aune\altaffilmark{4},
M.~Bautista\altaffilmark{5},
M.~Beilicke\altaffilmark{6},
W.~Benbow\altaffilmark{1},
D.~Boltuch\altaffilmark{2},
S.~M.~Bradbury\altaffilmark{7},
J.~H.~Buckley\altaffilmark{6},
V.~Bugaev\altaffilmark{6},
Y.~Butt\altaffilmark{8},
K.~Byrum\altaffilmark{9},
A.~Cannon\altaffilmark{10},
A.~Cesarini\altaffilmark{11},
Y.~C.~Chow\altaffilmark{3},
L.~Ciupik\altaffilmark{12},
P.~Cogan\altaffilmark{5},
W.~Cui\altaffilmark{13},
R.~Dickherber\altaffilmark{6},
T.~Ergin\altaffilmark{8},
S.~J.~Fegan\altaffilmark{3},
J.~P.~Finley\altaffilmark{13},
P.~Fortin\altaffilmark{14},
L.~Fortson\altaffilmark{12},
A.~Furniss\altaffilmark{4},
D.~Gall\altaffilmark{13},
G.~H.~Gillanders\altaffilmark{11},
E.~V.~Gotthelf\altaffilmark{15},
J.~Grube\altaffilmark{10},
R.~Guenette\altaffilmark{5},
G.~Gyuk\altaffilmark{12},
D.~Hanna\altaffilmark{5},
J.~Holder\altaffilmark{2},
D.~Horan\altaffilmark{16},
C.~M.~Hui\altaffilmark{17},
T.~B.~Humensky\altaffilmark{18},
P.~Kaaret\altaffilmark{19},
N.~Karlsson\altaffilmark{12},
M.~Kertzman\altaffilmark{20},
D.~Kieda\altaffilmark{17},
A.~Konopelko\altaffilmark{21},
H.~Krawczynski\altaffilmark{6},
F.~Krennrich\altaffilmark{22},
M.~J.~Lang\altaffilmark{11},
S.~LeBohec\altaffilmark{17},
G.~Maier\altaffilmark{5},
A.~McCann\altaffilmark{5},
M.~McCutcheon\altaffilmark{5},
J.~Millis\altaffilmark{23},
P.~Moriarty\altaffilmark{24},
R.~Mukherjee\altaffilmark{14},
R.~A.~Ong\altaffilmark{3},
A.~N.~Otte\altaffilmark{4},
D.~Pandel\altaffilmark{19},
J.~S.~Perkins\altaffilmark{1},
M.~Pohl\altaffilmark{22},
J.~Quinn\altaffilmark{10},
K.~Ragan\altaffilmark{5},
L.~C.~Reyes\altaffilmark{25},
P.~T.~Reynolds\altaffilmark{26},
E.~Roache\altaffilmark{1},
H.~J.~Rose\altaffilmark{7},
M.~Schroedter\altaffilmark{22},
G.~H.~Sembroski\altaffilmark{13},
A.~W.~Smith\altaffilmark{9},
D.~Steele\altaffilmark{12},
S.~P.~Swordy\altaffilmark{18},
M.~Theiling\altaffilmark{1},
J.~A.~Toner\altaffilmark{11},
V.~V.~Vassiliev\altaffilmark{3},
S.~Vincent\altaffilmark{17},
R.~G.~Wagner\altaffilmark{9},
S.~P.~Wakely\altaffilmark{18,**},
J.~E.~Ward\altaffilmark{10},
T.~C.~Weekes\altaffilmark{1},
A.~Weinstein\altaffilmark{3},
T.~Weisgarber\altaffilmark{18},
D.~A.~Williams\altaffilmark{4},
S.~Wissel\altaffilmark{18},
M.~Wood\altaffilmark{3},
B.~Zitzer\altaffilmark{13}
}

\altaffiltext{1}{Fred Lawrence Whipple Observatory, Harvard-Smithsonian Center for Astrophysics, Amado, AZ 85645, USA}
\altaffiltext{2}{Department of Physics and Astronomy and the Bartol Research Institute, University of Delaware, Newark, DE 19716, USA}
\altaffiltext{3}{Department of Physics and Astronomy, University of California, Los Angeles, CA 90095, USA}
\altaffiltext{4}{Santa Cruz Institute for Particle Physics and Department of Physics, University of California, Santa Cruz, CA 95064, USA}
\altaffiltext{5}{Physics Department, McGill University, Montreal, QC H3A 2T8, Canada}
\altaffiltext{6}{Department of Physics, Washington University, St. Louis, MO 63130, USA}
\altaffiltext{7}{School of Physics and Astronomy, University of Leeds, Leeds, LS2 9JT, UK}
\altaffiltext{8}{Harvard-Smithsonian Center for Astrophysics, 60 Garden Street, Cambridge, MA 02138, USA}
\altaffiltext{9}{Argonne National Laboratory, 9700 S. Cass Avenue, Argonne, IL 60439, USA}
\altaffiltext{10}{School of Physics, University College Dublin, Belfield, Dublin 4, Ireland}
\altaffiltext{11}{School of Physics, National University of Ireland, Galway, Ireland}
\altaffiltext{12}{Astronomy Department, Adler Planetarium and Astronomy Museum, Chicago, IL 60605, USA}
\altaffiltext{13}{Department of Physics, Purdue University, West Lafayette, IN 47907, USA }
\altaffiltext{14}{Department of Physics and Astronomy, Barnard College, Columbia University, NY 10027, USA}
\altaffiltext{15}{Columbia Astrophysics Laboratory, Columbia University, New York, NY 10027, USA}
\altaffiltext{16}{Laboratoire Leprince-Ringuet, Ecole Polytechnique, CNRS/IN2P3, F-91128 Palaiseau, France}
\altaffiltext{17}{Department of Physics and Astronomy, University of Utah, Salt Lake City, UT 84112, USA}
\altaffiltext{18}{Enrico Fermi Institute, University of Chicago, Chicago, IL 60637, USA}
\altaffiltext{19}{Department of Physics and Astronomy, University of Iowa, Van Allen Hall, Iowa City, IA 52242, USA}
\altaffiltext{20}{Department of Physics and Astronomy, DePauw University, Greencastle, IN 46135-0037, USA}
\altaffiltext{21}{Department of Physics, Pittsburg State University, 1701 South Broadway, Pittsburg, KS 66762, USA}
\altaffiltext{22}{Department of Physics and Astronomy, Iowa State University, Ames, IA 50011, USA}
\altaffiltext{23}{Department of Physics, Anderson University, 1100 East 5th Street, Anderson, IN 46012}
\altaffiltext{24}{Department of Life and Physical Sciences, Galway-Mayo Institute of Technology, Dublin Road, Galway, Ireland}
\altaffiltext{25}{Kavli Institute for Cosmological Physics, University of Chicago, Chicago, IL 60637, USA}
\altaffiltext{26}{Department of Applied Physics and Instrumentation, Cork Institute of Technology, Bishopstown, Cork, Ireland}

\altaffiltext{**}{Corresponding author: wakely@uchicago.edu}
\altaffiltext{*}{Corresponding author: ealiu@bartol.udel.edu}

\begin{abstract}

We report the detection of very-high-energy (VHE) gamma-ray emission from supernova remnant (SNR) G106.3+2.7. Observations performed in 2008 with the VERITAS atmospheric Cherenkov gamma-ray telescope resolve extended emission overlapping the elongated radio SNR. The 7.3$\sigma$ (pre-trials) detection has a full angular extent of roughly $0.6^\circ$ by $0.4^\circ$.  Most notably, the centroid of the VHE emission is centered near the peak of the coincident $^{12}$CO (J=1-0) emission, $0.4^\circ$ away from the pulsar PSR J2229+6114, situated at the northern end of the SNR. Evidently the current-epoch particles from the pulsar wind nebula are not participating in the gamma-ray production. The VHE energy spectrum measured with VERITAS is well characterized by a power-law $dN/dE = N_0(E/3\;TeV)^{-\Gamma}$ with a differential index of $\Gamma = 2.29 \pm 0.33_{stat} \pm 0.30_{sys}$ and a flux of $N_0 = (1.15 \pm 0.27_{stat} \pm 0.35_{sys}) \times 10^{-13}$ cm$^{-2}$s$^{-1}$TeV$^{-1}$. The integral flux above 1 TeV corresponds to $\sim 5$ percent of the steady Crab Nebula emission above the same energy.  We describe the observations and analysis of the object and briefly discuss the implications of the detection in a multiwavelength context.

\end{abstract}

\keywords{gamma rays: observations  --- ISM: individual (G106.3+2.7=VER J2227+608) --- pulsars: individual (J2229+6114) --- supernova remnants}

\section{Introduction}

The supernova remnant (SNR) G106.3+2.7 was first identified as a faint extended radio source in a northern Galactic Plane radio survey with the Dominion Radio Astrophysical Observatory (DRAO) by \citet{1990AAS...82..113J}.  Working with the earlier radio data of \citet{1980AAS...42..227K}, those authors determined that the object was a previously undetected supernova remnant, with a radio spectral index of $0.45$ and a flux density at 1~GHz of 6~Jy.  Subsequent work by \cite{2000AJ....120.3218P} confirmed the object as a supernova remnant, with an estimated age and distance of 1.3~Myr, and 12~kpc.  The object can be described as cometary in shape, with a compact radio-bright head to the north, and a dimmer, extended tail to the southwest.

Located at the northern edge of the remnant's head is the pulsar PSR~J2229+6114 and its associated boomerang-shaped radio and x-ray-emitting wind nebula, G106.6+2.9 \citep{2001ApJ...547..323H}.  The pulsar, which has a period of 51.6~ms, was discovered in a search of the error box of the EGRET source 3EG~J2227+6122 by \cite{2001ApJ...552L.125H} and is notable for being one of the most energetic pulsars known, with a spin-down power $\dot{E} = 2.2 \times 10^{37}$ erg~s$^{-1}$, despite having a low radio luminosity \citep{2006ApJ...638..225K}.

Based on polarization studies and velocity maps of HI and CO emission, \cite{2001ApJ...560..236K} associate the pulsar wind nebula and the supernova remnant with the same progenitor event and invoke a scenario of shock-wave evolution into a complex molecular environment to explain the unusual configuration of the constituent components.  These authors derive a distance to the remnant of 0.8~kpc, and adopt an age of 10~kyr, inferred from the pulsar data \citep{2001ApJ...552L.125H}.

At GeV energies, the EGRET source 3EG~J2227+6122, with relatively large error contours, is compatible with the pulsar position, as well as with the main bulk of the radio remnant \citep{1999ApJS..123...79H}.  The newly-released Large Area Telescope Bright Source List of the Fermi Gamma-ray Space Telescope (FGST) contains the object 0FGL J2229.0+6114, the position of which is consistent with the pulsar position \citep{2009arXiv0902.1559A}.  The AGILE source 1AGL J2231+6109 \citep{web:Agile} is also located nearby.

At TeV energies, the most stringent flux limits come from the MAGIC Collaboration \citep{EsterThesis}, who quote a point-source upper limit of $\sim10$\% of the Crab Nebula flux above 220~GeV at the pulsar position.  Less constraining point-source limits have also been presented by the Whipple \citep{2005ApJ...624..638F} and VERITAS \citep{2007arXiv0709.3975K} Collaborations.

Additionally, the Milagro Collaboration has recently reported a new analysis of their data for this region, resulting in the detection of gamma rays over a broad $\sim 1^\circ$ area encompassing both the pulsar position and the main bulk of the remnant \citep{2009arXiv0904.1018A}.  The detection corresponds to the location of their previously published ``source candidate C4" \citep{2007ApJ...664L..91A}, but lacks the angular resolution to provide a definitive association with a particular region within the SNR/pulsar complex. The statistical significance of the detection at the pulsar location is $6.6\sigma$ and the median energy of the detected gamma rays is 35~TeV.  The quoted differential flux is $(70.9 \pm 10.8_{stat} \pm 35.5_{sys}) \times 10^{-17}$ TeV$^{-1}$ cm$^{-2}$ s$^{-1}$, though this depends on assumptions about the source spectral shape.

\section{VERITAS Instrument \& Observations}

The VERITAS observatory comprises an array of four 12~m imaging atmospheric Cherenkov telescopes, located at Mt. Hopkins in southern Arizona (1268~m a.s.l., N~31$^\circ$40$^\prime$30$^{\prime\prime}$, W~110$^\circ$57$^\prime$07$^{\prime\prime}$).  The instrument, which features a 3.5$^\circ$ field-of-view, has a $5\sigma$ point source sensitivity of 1\% of the steady Crab Nebula flux in under 50 hours and an angular resolution of $\sim 0.1^\circ$, making it well-suited for the study of extended gamma-ray sources.  For a detailed description of the telescopes and their operation see \cite{2002APh....17..221W} or \cite{2006APh....25..391H}.  A recent review of the atmospheric Cherenkov technique can be found in \cite{2008RPPh...71i6901A}.

VERITAS observed G106.3+2.7 during three epochs.  The first was during November/December of 2006, with 2 telescopes, resulting in a point-source flux upper limit reported in \cite{2007arXiv0709.3975K}.  Observations with 4 telescopes continued in September-November of 2007, as a part of a larger survey for PWN emission from high-spin-powered pulsars in the northern sky \citep{Heidelberg08-Ester}.  After a hint of extended gamma-ray emission was seen in these data, follow-up observations were triggered in the 2008 observing season. In each epoch, the telescope tracking direction was centered on the position of the pulsar.

All observations were performed in the so-called ``wobble mode", where the telescopes point in a direction offset by some angle from the putative source position.  This is done to provide a simultaneous estimate of backgrounds with every run \citep{1994APh.....2..137F}.  The offset direction is sequentially varied from run to run between the four cardinal directions.  In the earlier observing epochs, small offset angles, suitable for point-source searches, were used.  Based on the traces of extended emission seen in the previous data, a 0.7 degree offset, more appropriate for larger sources, was adopted in the 2008 epoch.  Only data from this epoch, which also includes the deepest exposure, are included in this Letter.  After quality selection cuts to remove periods with poor weather or malfunctioning hardware, 111 observing runs remain, with a mean zenith angle of 34 degrees and a total deadtime-corrected exposure of 33.4 hours.

\section{Data Analysis and Results}\label{sec:anal}

\begin{figure*}[t]
\includegraphics[width=0.95\linewidth]{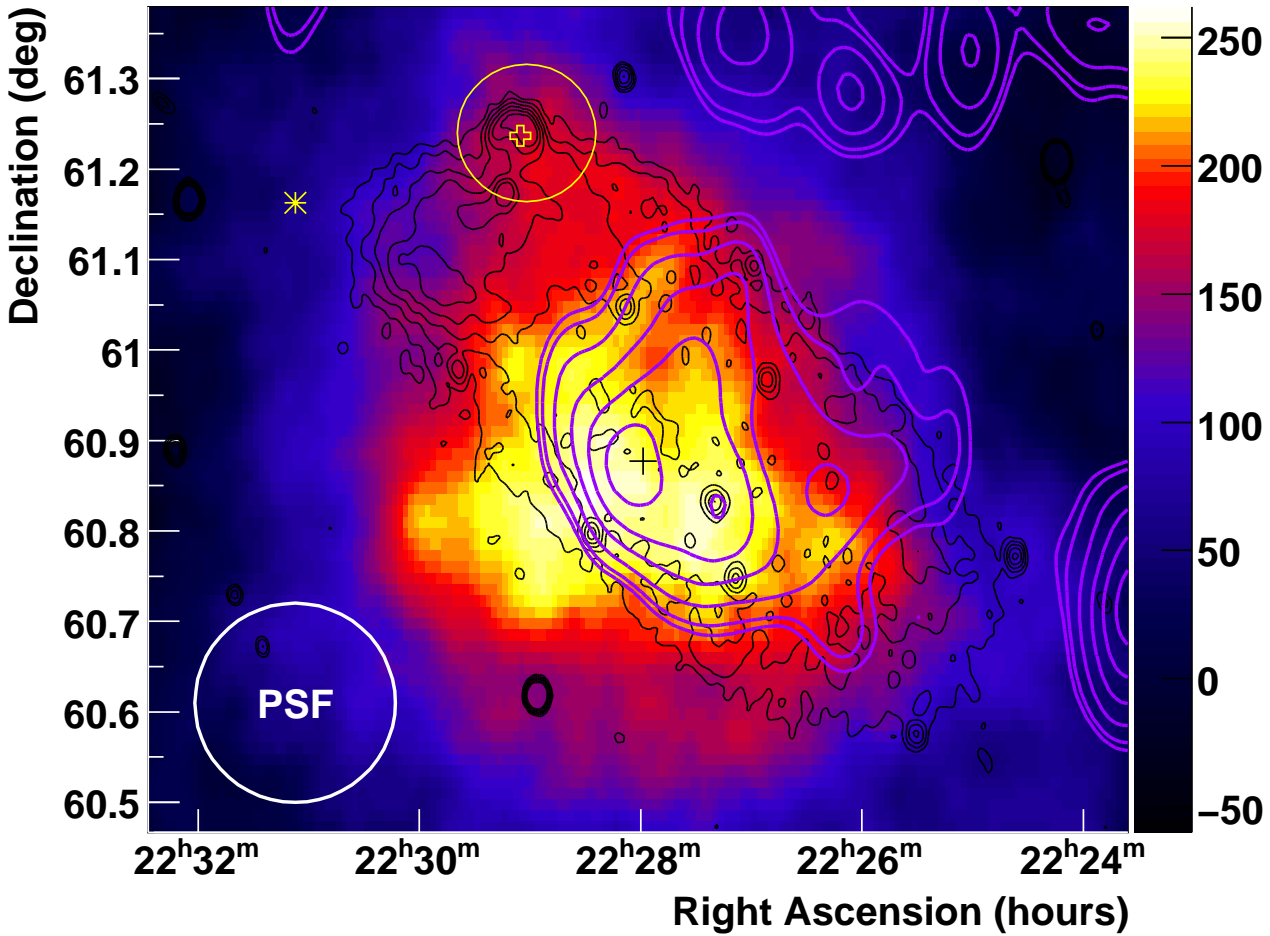} 
\caption{\label{fig:map} Sky map of TeV gamma-ray emission from G106.3+2.7, as measured by VERITAS.  The color scale indicates the number of excess gamma-ray events from the region, using a squared integration radius of $0.08\;\textrm{deg}^2$.  The centroid of the TeV emission is indicated with a thin black cross.  Overlaid are 1420 MHz radio contours from the DRAO Synthesis Telescope (thin black lines - \cite{2000A&AS..145..509L}) and $^{12}$CO emission (J=1-0) from the high-resolution FCRAO Survey (magenta lines - \cite{1998ApJS..115..241H}).  The CO velocity selection is discussed in the text.  The open yellow cross shows the location of pulsar PSR~J2229+6114 and the yellow star at the left is the AGILE bright source 1AGL~J2231+6109.  The yellow circle indicates the 95\% error contour for the Fermi source 0FGL~J2229.0+6114.  The circle labeled PSF represents the VERITAS gamma-ray point-spread function for this analysis (68\% containment).}
\end{figure*}

The data were analyzed using the VERITAS standard analysis package, described in \cite{2007arXiv0709.4233C} and \cite{2008ApJ...679.1427A}.  After gain corrections and cleaning, the air shower images in the individual telescopes are parameterized using the Hillas moment analysis \citep{1985ICRC....3..445H}.  Images which pass the pre-selection criteria (number of focal-plane phototubes in image $\geqslant 5$; distance from image centroid to camera center $< 1.43$ degrees; total image size $> 150$ photoelectrons) are then retained for full event reconstruction using the mean-scaled width and length parameters (see \textit{e.g.}, \cite{Kono-MSW,2007arXiv0709.4006D}).  Events where only a single telescope image survives cuts are rejected, as are events where the only surviving images are from telescopes T1 and T4, which, with a separation of 35~m, are too close together for reliable event reconstruction.  The gamma/hadron cuts were based on an optimization using a large sample of Crab Nebula data, scaled to a flux level of 3\%.  The energy threshold of the analysis is 630 GeV.

Background estimation for the spectral reconstruction was performed using the reflected-region model \citep{2001A&A...370..112A}.  For sky map construction (see Figure \ref{fig:map}), the ring-background model was employed (for a discussion, see \cite{2007A&A...466.1219B} and references therein).  In the
search for extended emission, a squared angular integration radius of $0.08\;\textrm{deg}^2$, well-suited for the investigation of larger sources, was used.  Additionally, a smaller radius was used in a point source search.  The statistical significance of any excess event counts were evaluated using Equation (17) from \cite{1983ApJ...272..317L} and the results were confirmed using an independent internal analysis chain. There are no bright stars in the field of view to complicate observations or analysis.

Using the extended emission cuts, a pre-trials maximum significance of $7.3\sigma$ was observed.  The location of this peak significance is $\sim 0.4$ degrees to the south of the position of the pulsar PSR~J2229+6114.  Because there is no \textit{a priori} test point at this location, we apply a conservative trials penalty determined by tiling the entire area defined by the remnant's outer radio contour with 0.04$^\circ$ square search bins.  An additional factor associated with the small set of search cuts is also incorporated.  With these included, the post-trials significance of the peak emission point is $6.0\sigma$.  The post-trials significance of the excess seen at the nominal pulsar location is $3.6\sigma$.

\subsection{Morphology}
Inspection of Figure \ref{fig:map} shows that the angular extent of the TeV emission region exceeds the size of the $0.11^\circ$ VERITAS point-spread-function (PSF; the radius that contains 68\% of
the events coming from a point source).  While the shape of the extension in this map is apparently more complex than a simple Gaussian in nature, a statistically acceptable fit can nevertheless be obtained by fitting a 2-dimensional Gaussian to the uncorrelated acceptance-corrected map of excess event counts, binned in $0.05^\circ$ bins.  The result of this analysis shows that the TeV emission, after accounting for the PSF of the instrument, can be characterized by a 1$\sigma$ angular extent of $0.27^\circ \pm 0.05^\circ$ along the major axis, and $0.18^\circ \pm 0.03^\circ$ along the minor axis; the orientation angle is $22^\circ$ east of north.  The centroid of the fit lies at $22^{\rm h} \ 27^{\rm m} \ 59^{\rm s},\ +60^{\circ} \ 52^{\prime} \ 37^{\prime\prime}$ (J2000) and hence we assign the identifier VER~J2227+608.  The statistical uncertainty in the centroid position is $0.07^\circ$ in RA and $0.04^\circ$ in declination, with an additional combined systematic uncertainty of $0.07^\circ$.

\subsection{Spectrum}
The spectrum of the source was determined by integrating over a circular region of radius $0.32^\circ$, positioned to encompass the majority of the observed emission.  The resulting spectrum, which is shown in Figure \ref{fig:spectrum}, can be fitted ($\chi^2$=0.76;ndf=3) with a simple power-law of the form $dN/dE = N_0 (E/3\;TeV)^{-\Gamma}$, with a differential flux constant of $N_0 = (1.15 \pm 0.27_{stat} \pm 0.35_{sys}) \times 10^{-13}$ cm$^{-2}$s$^{-1}$TeV$^{-1}$ and an index of $\Gamma = 2.29 \pm 0.33_{stat} \pm 0.30_{sys}$.  The total flux integrated over the remnant for energies above 1~TeV is $(1.11 \pm 0.25_{stat} \pm 0.28_{sys}) \times 10^{-12}$ cm$^{-2}$ s$^{-1}$, which corresponds to $\sim 5\%$ of the steady Crab Nebula flux.  Also shown in this figure is the Milagro flux point from \cite{2009arXiv0904.1018A}.  The energy resolution of this analysis is better than $\sim 20\%$ for energies above 1 TeV.

\section{Discussion}

Figure \ref{fig:map} shows the TeV gamma-ray image for G106.3+2.7.  The color scale indicates the number of excess gamma-ray candidate events, as measured by VERITAS.  Radio continuum emission at 1420 MHz measured by the Dominion Radio Astrophysical Observatory (DRAO) Synthesis Telescope \citep{2000A&AS..145..509L}, and supplied through the Canadian Galactic Plane Survey \citep{2003AJ....125.3145T}, is indicated with thin black contours.  A high-resolution map of 115 GHz line emission associated with the molecular $^{12}$CO~(J=1-0) transition, from the Five College Radio Astronomy Observatory (FCRAO) is shown in magenta contours \citep{1998ApJS..115..241H}.  Following the analysis of \cite{2001ApJ...560..236K}, the CO velocities in the map are restricted to between -4~km~s$^{-1}$ and -6~km~s$^{-1}$.  This slice encompasses the majority of the cloud emission overlapping the radio remnant.  Note that the CO contours have been smoothed with a $\sim 3^\prime$  Gaussian kernel.  The open yellow cross marks the location of PSR J2229+6114; its wind nebula, the ``Boomerang" ($\sim 0.08^\circ$ in diameter), can be seen in the radio contours around it.  The yellow star at the left is the AGILE bright source 1AGL J2231+6109 and the 95\% FGST confidence contour for the source 0FGL~J2229.0+6114 is shown as the yellow circle centered on the pulsar location.

As can be seen from Figure \ref{fig:map}, the TeV emission appears correlated with the position of the radio contours of the main body of the remnant, reaching maximum flux $\sim 0.4^\circ$ away from the pulsar.  Indeed, the region of greatest intensity is apparently coincident with the location of the molecular cloud emission.  This coincidence is reminiscent of behavior seen in similar objects (see, \textit{e.g.}, \cite{2009ApJ...698L.133A, 2008A&A...483..509A, 2008A&A...481..401A}) and is supportive of theories of hadronically-induced (that is, $p+p \rightarrow \pi^0 \rightarrow \gamma \gamma$) emission \citep{1994A&A...285..645A,1996A&A...309..917A}.  The detection of OH maser emission, a signpost of shock wave/cloud interactions \citep{1976ApJ...203..124E,2002Sci...296.2350W}, would strengthen this claim, but no such observations have been reported.

\begin{figure}[t]
\includegraphics[width=0.9\linewidth]{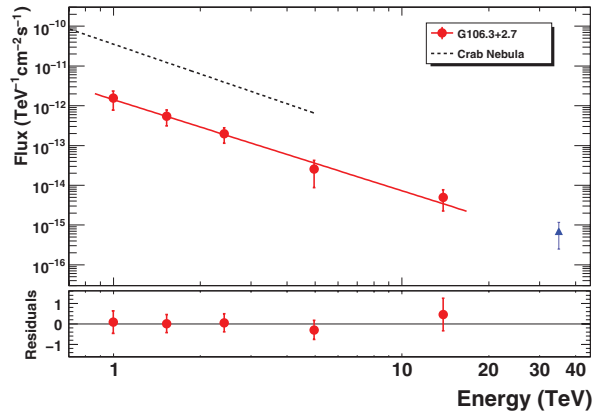}
\caption{\label{fig:spectrum} Differential energy spectrum of G106.3+2.7 as measured by VERITAS.  The error bars represent $1\sigma$ statistical errors only.  The solid line shows the results of a pure power-law fit and the dotted line shows the flux of the Crab Nebula for comparison.  Also shown (triangle) is the Milagro flux point from \cite{2009arXiv0904.1018A}, plotted with summed systematic and statistical errors.  The lower panel shows the residuals of the VERITAS data to the power-law fit.  The details of the analysis and the fit are discussed in the text.}
\end{figure}

A leptonic scenario, wherein gamma rays are produced by the inverse-Compton scattering of a local photon pool (for instance, synchrotron-produced photons, cosmic microwave background (CMB) photons, or infrared photons) by high-energy electrons, produced either by the SNR, or the PWN, is another possibility.  \cite{2005JPhG...31.1465B} have applied such a model to the Boomerang PWN and made predictions for the expected gamma-ray flux.  In this model, leptons can obtain high energies through resonant scattering off Alfven waves generated by heavy nuclei, which, in turn, have been accelerated directly by the pulsar or PWN.  Assuming only photon targets of synchrotron and CMB origin, they obtain an integral flux prediction which corresponds to (in the absence of any cutoff) $1.6 \times 10^{-12}$ photons cm$^{-2}$ s$^{-1}$ above 1 TeV.  This prediction is consistent within statistical and systematic errors with the present measurement.  The lack of a clear spatial coincidence between the PWN and the TeV emission is not unique (\textit{e.g.}, \cite{2006A&A...448L..43A, 2006A&A...456..245A}) and can be explained in an older ($\tau \gtrsim 10^4$ yrs) remnant such as this because the emission is dominated by the scattering of CMB photons by lower-energy leptons.  These leptons may be displaced from the present pulsar position due either to pulsar motion or to interactions between the PWN and SNR \citep{2003A&A...405..689B,2008MNRAS.385.1105B}.  A possible gamma-ray component due to hadrons trapped in nearby molecular clouds could also contribute ({\it e.g.}, \cite{2008MNRAS.385.1105B}).  Finally, we note that our measured spectrum, as shown in Figure \ref{fig:spectrum}, if extrapolated to 35~TeV, is consistent within errors with the flux reported by the Milagro group in \cite{2009arXiv0904.1018A}.  Given the uncertainty in the Milagro source position, it isn't possible to unambiguously associate the two sources.  However, if there is, in fact, a single mechanism at work in producing the emission seen by both instruments, then this could weaken the case for a leptonic scenario, given that the combined spectrum lacks the curvature expected from a typical inverse Compton model at these energies (see, \textit{e.g.}, \cite{2008arXiv0803.0116D}).

\section{Conclusion}
The VERITAS observatory has made a high-confidence detection of TeV gamma rays from the supernova remnant G106.3+2.7.  This gamma-ray emission is significantly extended, and seems well-correlated with the radio extent of the remnant, in particular, with the location of known nearby molecular clouds.  The emission seen at the location of the associated pulsar PSR J2229+6114 is non-zero, but only of marginal statistical significance.

The total flux from the remnant above 1 TeV is about $\sim 5\%$ of the Crab Nebula, and can be described by a power-law in energy with differential index 2.29.  With the information available, it is not yet possible to make a firm determination of whether the TeV gamma-ray emission is of leptonic or hadronic origin, though a solid association with the Milagro source could favor the latter.  With additional exposure and a refined observation strategy (\textit{i.e.}, wobbling around the location of the emission peak rather than the pulsar) a more in-depth study of the object will be achievable, including an investigation of possible energy dependence in the source morphology.  Such an effort will benefit from the enhanced sensitivity and angular resolution of the reconfigured VERITAS array, in which the T1 telescope (see Section \ref{sec:anal}) is relocated to obtain a more optimal baseline.  This upgrade will be completed in autumn, 2009.

\acknowledgements
This research is supported by grants from the U.S. Department of Energy, the U.S. National Science Foundation and the Smithsonian Institution, by the Natural Sciences and Engineering Research Council (NSERC) in Canada, by Science Foundation Ireland and by the Science and Technology Facilities Council in the UK. The research presented in this paper has used data from the Canadian Galactic Plane Survey, a Canadian project with international partners, supported by NSERC. We acknowledge the work of the technical support staff at the FLWO, of BtWD, and of the collaborating institutions in the construction and operation of the instrument.

{\it Facilities:} \facility{FLWO:VERITAS}.

\bibliographystyle{hapj}
\bibliography{Boomerang}

\end{document}